\documentclass[twocolumn,showpacs,preprintnumbers,amsmath,amssymb]{revtex4}

\usepackage{graphicx}
\usepackage{bm}
\usepackage{dcolumn}     

\begin{document}
\title{Electromagnetic transition from the 4$^+$ to 2$^+$ resonance in $^8$Be
measured via the radiative capture in $^4$He+$^4$He} 
\author{V. M. Datar$^{1,2}$} 
\author{D. R. Chakrabarty$^1$} 
\author{Suresh Kumar$^{1,2}$}
\author{V. Nanal$^3$} 
\author{S. Pastore$^4$}
\author{R. B. Wiringa$^5$}
\author{S. P. Behera$^1$}
\author{A. Chatterjee$^1$}
\author{D. Jenkins$^6$} 
\author{C. J. Lister$^5$} 
\author{E. T. Mirgule$^1$}
\author{A. Mitra$^1$}
\author{R. G. Pillay$^3$} 
\author{K. Ramachandran$^1$}
\author{O. J. Roberts$^6$} 
\author{P. C. Rout$^{1,2}$}
\author{A. Shrivastava$^1$}
\author{P. Sugathan$^7$} 
\affiliation{$^1$Nuclear Physics Division, Bhabha Atomic Research Centre,
Mumbai 400 085, India}
\affiliation{$^2$Homi Bhabha National Institute, Anushaktinagar, Mumbai 400 094, India}
\affiliation{$^3$Tata Institute of Fundamental Research, Mumbai 400 005, India}
\affiliation{$^4$Department of Physics and Astronomy, University of South Carolina, Columbia, SC 29208, USA}
\affiliation{$^5$Physics Division, Argonne National Laboratory, Argonne, IL 60439, USA}
\affiliation{$^6$Department of Physics, University of York, Heslington, York, Y010 5DD, UK}
\affiliation{$^7$Inter University Accelerator Centre, New Delhi-110064, India}

\date{\today}

\begin{abstract}
An earlier measurement on the 4$^+$ to 2$^+$ radiative transition in $^8$Be 
provided the first electromagnetic signature of its dumbbell-like shape. 
However, the large uncertainty in the measured cross section 
does not allow a stringent test of nuclear structure models. The present paper reports 
a more elaborate and precise measurement for this transition,
via the radiative capture in the $^4$He+$^4$He reaction, improving the
accuracy by about a factor of three. The {\it ab initio} calculations of the
radiative transition strength with improved three-nucleon forces are also presented.
The experimental results are compared with the
predictions of the alpha cluster model and {\it ab initio} calculations. 
\end{abstract}
\maketitle

The nucleus $^8$Be is a classic example of the occurrence of alpha clustering~\cite{bm2} in nuclei. 
Its formation from two alpha particles provides an intermediate step 
in the synthesis of $^{12}$C~\cite{hoyle} from the fusion of three alpha particles inside the stars. 
The nucleus is also the stepping stone to understand alpha-clustering in heavier self-conjugate 4n nuclei.
The dumbbell-shaped nucleus exhibits rotational states manifested as resonances in the 
alpha-alpha scattering system. 
The electromagnetic transition between the excited resonant states in $^8$Be, with spin-parities 
of 4$^+$ and 2$^+$, 
was reported earlier~\cite{datarPRL} in order to provide a test for its 
alpha cluster structure. The measurements were made at two beam energies, on and off the 4$^+$ 
resonance, by detecting the transition gamma rays
in coincidence with the two alpha particles arising from the
decay of the  2$^+$ final state. However, 
the measured cross section (with an uncertainty 
of $\sim$33\%) and the inferred reduced 
electromagnetic transition rate were not precise enough  to provide a stringent test for
various models like the cluster model~\cite{langanke} and {\it ab initio} 
quantum Monte Carlo model~\cite{WPCP00}. The uncertainty arose mainly due to the large background 
of 4.44~MeV gamma rays 
originating from the interaction of the incident beam with the window of the chamber 
holding the helium gas target. The present work, using essentially the same method, is aimed at a more 
accurate measurement 
and also at more beam energies straddling the 4$^+$ resonance.
The essential aspects in this improved measurement are a better pixelisation of the alpha particle detectors,
a more efficient and segmented gamma ray detector and a better shielding of the gamma rays from 
the beam-window interaction mentioned above.

\begin{figure}
\includegraphics[scale=0.25]{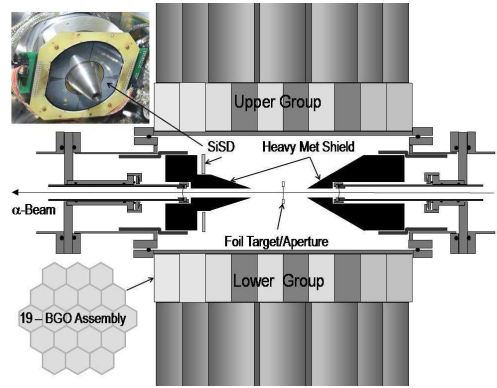} 
\caption{(Color online) Schematic of experimental setup.}
\end{figure}

The experiment was carried out using beams of $^4$He from the BARC-TIFR Pelletron 
Linac Facility at TIFR, Mumbai at energies of 19$-$29 MeV. The beam 
current was about 1~pnA on the target. The schematic of the experimental setup is shown in Fig.~1.
The $\gamma$-rays were detected in a BGO detector array
with a photopeak efficiency of about 23\% at E$_\gamma$=8 MeV. The 
array consisted of 38 hexagonal cross section detectors, of length 76~mm and 
a face to face distance of 56 and 58~mm (in two groups), encased in thin aluminum housing. These were
mounted in close packed groups of 19 each placed at $\sim$70~mm above and below the target. 
Alpha particles were detected in a 500~$\mu$m thick, annular, and double sided silicon strip 
detector (SiSD), with 2$\times$16~$\theta$ rings (in left and right halves) 
and 16~$\phi$ sectors~\cite{sisd} with separate readouts. 
The active portion of the detector had an 
inner diameter of 48~mm and an outer diameter of 96~mm. The gap between the adjacent rings was 0.1~mm 
while that between adjacent sectors was 0.2~mm. 
The left and right halves of the $\theta$ side were separated by 0.4~mm.

\begin{figure}
\includegraphics[scale=0.50]{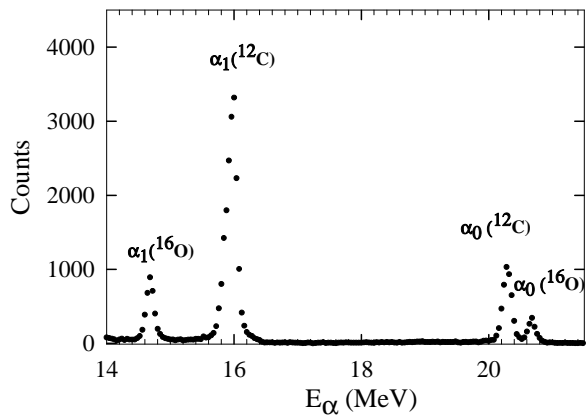}
\caption{Alpha particle spectra from the scattering on the mylar target at E$_{\rm beam}$=22.4 MeV
for a particular ring of the SiSD.
Elastic and inelastic peaks (from the first excited states) from the 
$^{12}$C and $^{16}$O targets are indicated.} 
\end{figure}
A chamber was designed to mount the strip detector at $\sim$70 mm from its centre 
in the forward direction and to hold the target helium gas (purity $>$99.9\%) at  $\sim$0.8~bar pressure.
The gas was isolated from the beam line vacuum using $\sim$1~mg/cm$^2$ Kapton foils 
at the entry and the exit. 
The helium 
gas was filled using a gas handling setup similar to that used 
earlier~\cite{datarPRL}. Conical heavymet shields surrounded the Kapton windows 
in order to shield the BGO detector array from the copious 4.44~MeV $\gamma$-rays produced in the 
excitation of $^{12}$C in the windows. 
The chamber had the provision to mount a ladder for holding a thick aluminum aperture plate
with a hole of 24~mm diameter as well as thin mylar and carbon foils. The aperture plate, when placed 
at the centre of the target chamber, shielded the $\alpha$-particles scattered 
from the Kapton entrance window and
limited the effective beam-target interaction zone seen 
by the SiSD. The aperture diameter and the SiSD distance were decided on the basis of a 
Monte Carlo simulation~\cite{datarNPS} to get a reasonable efficiency for 
the detection of two $\alpha$-particles following the radiative capture and subsequent decay 
of the final state in $^8$Be. The typical effective target length was 
about 20~mm and the efficiency for the 2-$\alpha$ detection from the final state 
was about 35\%, after including the effect of the various dead zones in the SiSD. 

The energy and timing signals of the SiSD were generated from each of the 32 $\theta$-rings (divided into 
two groups of the left and right halves) and 16 $\phi$-sectors.
The energy signals were sent to voltage sensitive analog-to-digital converters. 
Each timing signal 
was fanned out into two paths one being used to generate the overall particle event 
trigger for left rings, right rings and the sectors using a logical OR condition among 
the corresponding signals. In the other path
the signals were fed to time to digital converters~(TDCs) for measuring timing with 
respect to the $\gamma$-ray detector
array. The anode signal from the photomultiplier of each of the 38 BGO 
detectors was also fanned out for measuring the energy deposited 
by a charge-to-digital 
converter and for the timing measurement with respect to the SiSD using TDCs. 
A logical OR condition among the 38 signals produced
the $\gamma$-ray event trigger. The grand event trigger was generated by 
requiring a fast coincidence between 
event triggers from the left and right halves of the SiSD $\theta$-rings and that 
from the BGO-detector array. 
The data were collected in an event by event mode using a CAMAC based data 
acquisition~(DAQ) system~\cite{lamps}. A 10~ Hz pulser signal was fanned out 
and given to the test input of the three SiSD preamplifiers for estimating the dead 
time of the DAQ system.

The energy calibration of the SiSD detector was done using  
elastic and inelastic scattering of $\alpha$-particles on $^{12}$C and $^{16}$O 
using thin carbon and mylar targets. A typical
$\alpha$-particle energy spectrum is shown in Fig. 2. The energy calibration was 
performed over the 256 (16$\times$16) $\theta-\phi$ pixels. The 4.44~MeV 
and 6.13~MeV $\gamma$-rays from excited states in $^{12}$C and $^{16}$O 
populated through the inelastic $\alpha$-particle scattering 
were used to calibrate the BGO detectors. These measurements were made periodically 
throughout the experiment in order to track the possible change in the calibrations of the 
$\alpha$-particle and $\gamma$-ray detectors. A stability within $\sim$1\% was witnessed over 
the period of the experiment.  

The data were collected at four beam energies of 19.2, 22.4, 24.7, 28.9 MeV,
spanning the 4$^+$ resonance in $^8$Be, for the integrated beam charges 
of 81, 90, 125 and 58 pnC, respectively. 
The data was analyzed to extract
the events corresponding to the $\gamma$-ray transition to the 2$^+$-final state
in $^8$Be and the subsequent 2-$\alpha$ decay of the final state.
The first condition imposed was the prompt concidence among the $\gamma$-ray detector,
at least one of the left rings and at least one of the right rings. 
This ensured a prompt coincidence also between the left and the right halves of the SiSD. 
The sector timing was also demanded to be in prompt coincidence
with the $\gamma$-ray detector with two opposite sectors being simultaneously in coincidence.
These conditions emphasized on the required events because the two $\alpha$-particles from the decay
of the final state are emitted at the azimuthal angles 
differing by $\sim$~180$^\circ$, neglecting the small momentum kick due to the 
transition $\gamma$-ray. Fig.~3 shows an 
\begin{figure}
\includegraphics[scale=0.45]{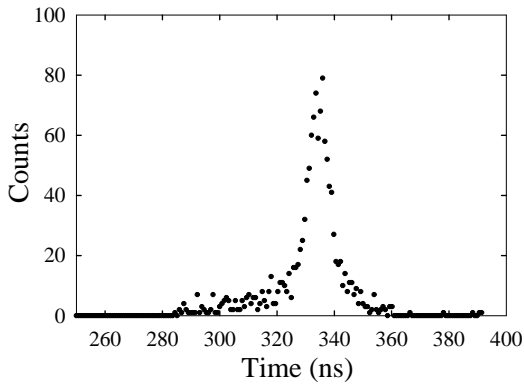}
\caption{Example of the time spectrum between the BGO detector array and the SiSD detector rings at 
E$_{\rm beam}$=22.4 MeV.}
\end{figure}
example of the time spectrum between the BGO detector array and the SiSD 
at a beam energy of 22.4~MeV showing the prompt time peak.
The hit-multiplicities of the left and the right rings were constrained to one for each and the energy
deposited in the left and right halves~(E$_L$ and E$_R$) were constructed from 
the energy calibrations of the corresponding rings. 
For getting the $\gamma$-ray energy E$_\gamma$, 
the BGO detector with the highest $\gamma$-ray energy was taken to be the primary detector.
The energies deposited in the neighbouring detectors, which were also in prompt coincidence and contained
the leaked shower energy, were added to that of
the primary detector for each event. An event by event reconstruction of the  
total $\alpha$-particle energy, E$_{sum}~=~$E$_L$~+~E$_R$ was made with conditions of
(a) both E$_L$ and E$_R$ being within a lower and a upper limit ($\sim$1$-$13 MeV) and
(b) the reconstructed total energy and the total momentum of the two $\alpha$-particles being within 
a proper two dimensional gate. These conditions were guided by the Monte Carlo simulations described below.  

Fig.4 shows a two dimensional 
plot of E$_{sum}$ vs E$_\gamma$ at the beam energy of 22.4~MeV. A band of events 
around an E$_{sum}$ of 13~MeV and E$_\gamma$ of 8~MeV 
can be clearly identified. These events arise from the 
radiative capture of the two $\alpha$-particles to the 2$^+$ resonance in $^8$Be.
A one dimensional spectrum of E$_{tot}$~=~E$_{sum}$~+~E$_\gamma$ is generated 
by putting one dimensional gates on E$_{sum}$ of 8.8$-$15.0 MeV and E$_\gamma$ of 3.4$-$10.5~MeV
as suggested by simulation results. Similar E$_{tot}$ spectra were generated at other beam 
energies by putting appropriate gates on these quantities. Fig.5 shows the E$_{tot}$ spectra 
at all the four beam energies. The peaks in the spectra (not apparent at the highest beam energy
which is beyond the extent of the 4$^+$-resonance), corresponding to the 
$\gamma$-ray transition between the resonances, were used in the calculation of the 
capture cross sections.   

The extraction of radiative capture cross sections requires a simulation 
of the experimental set up using a Monte Carlo code. Such a simulation was 
done in two parts -  one for the detection of the 
two $\alpha$-particles and the other for the response of the BGO 
\begin{figure}
\includegraphics[scale=0.55]{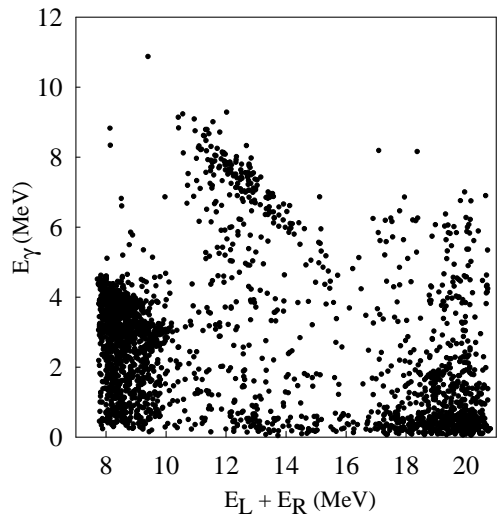}
\caption{Two dimensional spectrum between E$_{sum}$ =E$_L$+E$_R$ (see text)
and E$_\gamma$, generated with proper cuts as discussed in the text, at 
E$_{\rm beam}$=22.4 MeV.}
\end{figure}
array to the incident $\gamma$-rays. The simulation for $\alpha$-particle detection 
took into account the extended gas target, the aperture, the angular 
\begin{figure}
\includegraphics[scale=0.50]{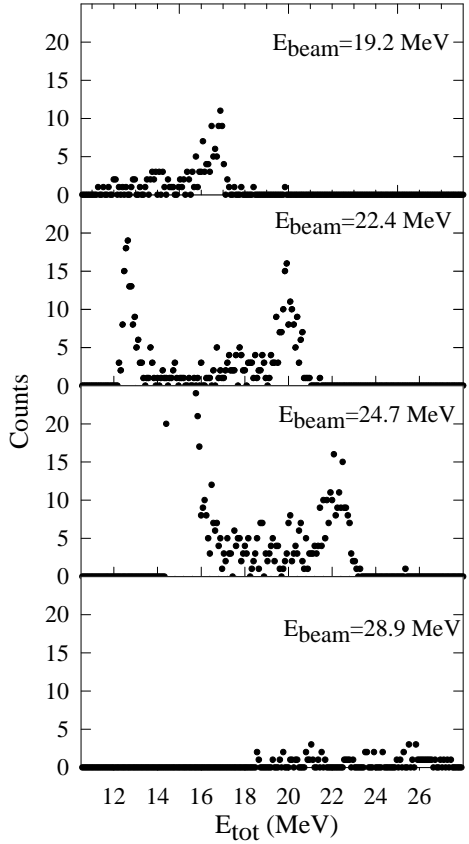}
\caption{One dimensional spectra of E$_{tot}$ =E$_{sum}$+ E$_\gamma$
 generated with proper cuts at four beam energies.}
\end{figure}
distribution of the $\alpha$-particles emitted after the $\gamma$-decay 
and the geometry of the SiSD. The energy losses of the beam and decay 
$\alpha$-particles, and the angular straggling, were calculated 
for each event using the SRIM code~\cite{srim}. 
The inefficiencies of the SiSD due to gaps between 
rings and sectors were included. The 
$\gamma$-ray response was simulated using GEANT3~\cite{geant3} with the 
angular distribution from the aligned 4$^+$ to the 2$^+$ final state included. 
For each event, the $\gamma$-ray energy was Doppler corrected. The 
simulated event by event data was written in a file for analysis by the 
same program that was used to sort the actual data.  
The simulated data were sorted to create the E$_{tot}$ spectrum with the 
same conditions as used in the case of the actual data. Starting from the N$_0$ events corresponding to the 
4$^+$ to 2$^+$ transition and the subsequent 2-$\alpha$ decay, the counts N in the same
peak regions as shown in Fig.5 were calculated from the simulated spectra to get the overall detection efficiency
(N/N$_0$) of the experimental set up. The simulation also provided the effective target thickness.
The capture cross sections were extracted using the integrated beam charge, the target thickness 
and the detection efficiency. 
The effective $\alpha$-particle energy $<$E$_\alpha>$
at each beam energy was also extracted from the simulation
after knowing the interaction region and the energy loss of the incident beam in the entrance window and
in the target gas up to the interaction region. The spread in the effective energy due to
the finite extent of the interaction region was less than 0.14~MeV.  
The extracted cross sections at the four beam energies and the corresponding effective $\alpha$-particle energies
are tabulated in Table~I. 

\begin{table}[!h]
\caption{Effective $\alpha$-particle energies $<$E$_\alpha>$ and the radiative capture cross sections 
$\sigma_\gamma$ extracted from the data at four beam energies (E$_{beam}$). }
\begin{ruledtabular}
\begin{tabular}{ccccc}
E$_{beam}$ & $<$E$_\alpha>$ & $\sigma_\gamma$ \\
(MeV) & (MeV)& (nb)\\
\hline
19.2&18.44&102$\pm$12 \\
22.4&21.80&149$\pm$16 \\
24.7&24.08&131$\pm$13 \\
28.9&28.40&$<$15 \\
\end{tabular}
\end{ruledtabular}
\end{table}

The extracted cross sections are plotted against the effective $\alpha$-particle energies in Fig.6. 
The cross section at the resonance energy is consistent with the earlier 
measurement~\cite{datarPRL}, but with an error of $\sim$10\% as compared 
to the earlier 33\%. 
\begin{figure}
\includegraphics[scale=0.55]{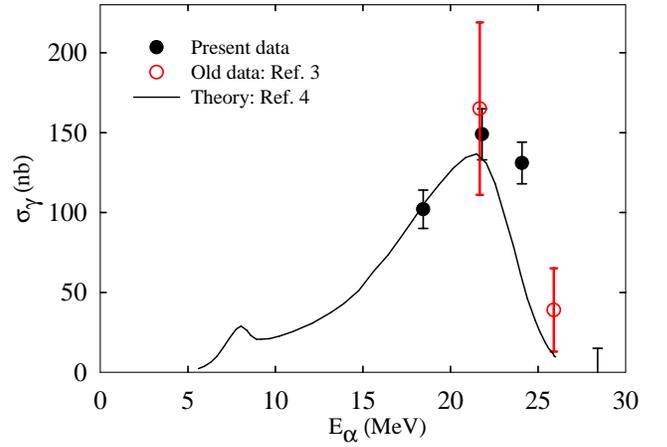}
\caption{(Color online) Extracted capture cross sections plotted against the effective
$\alpha$-particle energy (see text). The last point indicates the upper bound of the cross section. 
The continuous line shows the result of  a model calculation.}
\end{figure}
Fig.6 also shows the calculated cross sections from the cluster model calculation 
of~\cite{langanke}. The contribution from the partial waves of $l$=0, 2, 4 are added incoherently 
in this plot. The comparison with experimental data is
good in the rising part of the cross section profile but deviates at higher energies. 
Whether a different choice of the $\alpha-\alpha$ 
potential along with a coherent summing over the various partial waves will improve the
comparison remains to be seen. It may 
be mentioned that there is some ambiguity in the choice of the 
potentials giving similar values for the energies and widths of the resonant states of $^8$Be. 
\begin{figure}[!htb]
\includegraphics[height=0.45\textheight,keepaspectratio=true]{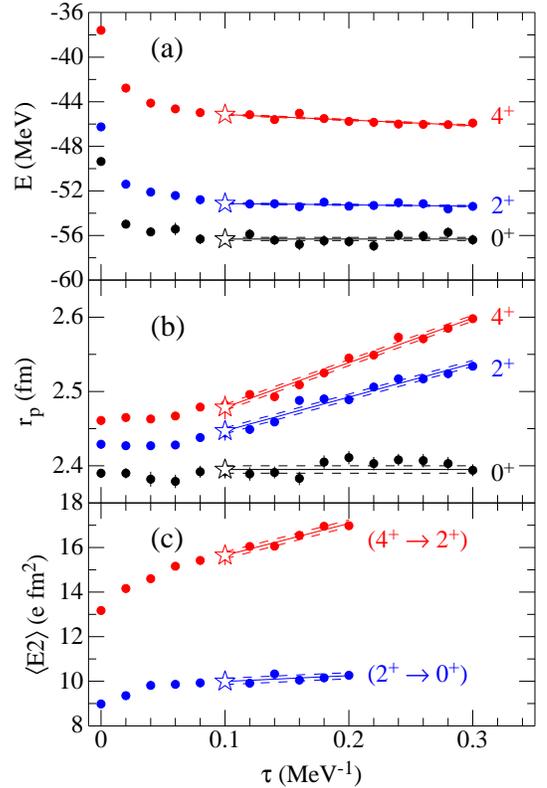}
\caption{(Color online) GFMC propagation with imaginary time $\tau$ of the 
(a) energy, (b) point proton radius, and (c) $E2$ matrix element; open
stars denote the values extracted from the calculation.}
\label{fig:gfmc}
\end{figure}

{\it Ab initio} calculations of the radiative transition strengths in $^8$Be,
using realistic two- and three-nucleon interactions, were first reported in~\cite{WPCP00}.
These variational Monte Carlo (VMC) calculations of the electric quadrupole
moment $Q$ and $B(E2)$ values indicated that the low-lying spectrum of $^8$Be
is well-described by the rotation of a common deformed two-$\alpha$ structure.
Recently it has become possible to evaluate electroweak transitions with the
more accurate Green's function Monte Carlo (GFMC) method~\cite{PPW07},
while improvements in the three-nucleon forces have also been made.
We report here new GFMC calculations using the Argonne $v_{18}$ (AV18)
two-nucleon~\cite{WSS95} and Illinois-7 (IL7) three-nucleon~\cite{P08}
potentials, which give a very nice reproduction of the energy
spectrum and other properties of light nuclei in this mass range~\cite{PPSW13}.

An initial VMC calculation is made to generate a starting wave function,
which the GFMC method then systematically improves upon by a propagation in
imaginary time $\tau$.
The $^8$Be~$2^+$ and $4^+$ excited states are particularly challenging
because they tend to break up into two separate $\alpha$-particles as $\tau$ increases.
Figure \ref{fig:gfmc} shows the propagation with imaginary time of the 
energies, radii, and $E2$ matrix elements.
In Fig.~\ref{fig:gfmc}(a), the energies of the states are seen to drop 
rapidly from the initial VMC energies at $\tau = 0$.
The $0^+$ ground state energy stabilizes and is well-fit by a constant 
averaged over $\tau = 0.1-0.3$ MeV$^{-1}$.
The $2^+$ state shows a very subtle decrease over the same range, while
the $4^+$ state drifts significantly lower; the energies quoted below are 
obtained from a linear fit using the value at $\tau = 0.1$ MeV$^{-1}$, with 
the Monte Carlo statistical error augmented by the range of values from
$\tau = 0.08-0.12$ MeV$^{-1}$.
This choice of $\tau$ should encompass the bulk of the improvement in the
wave functions provided by the GFMC algorithm, before the tendency to
dissolve into two $\alpha$-particles sets in.

This tendency to dissolution is seen more strongly in the evolution of the
point proton radii shown in Fig.~\ref{fig:gfmc}(b).
The $0^+$ ground state radius is flat as a function of $\tau$, while the
$2^+$ and $4^+$ states both increase steadily from about $\tau=0.1$ MeV$^{-1}$.
The associated electric quadrupole moments, which are not shown, also increase
steadily.
Finally, the $E2$ matrix elements, shown in Fig.~\ref{fig:gfmc}(c), also 
increase with $\tau$, the effect being particularly pronounced with the 
$(4^+ \rightarrow 2^+)$ transition.

Results for the energies $E$, point proton radii $r_p$, electric quadrupole 
moments $Q$, and $B(E2)$ transition strengths are given in Table \ref{tab:gfmc}.
The absolute energies of the states are in excellent agreement with experiment.
The quadrupole moments and $B(E2)$ values are consistent with an intrinsic
quadrupole moment $Q_0$ of $32 \pm 1$ fm$^2$, which is about $\sim 20\%$
bigger than the original VMC calculation of~\cite{WPCP00}.

\begin{table}[!h]
\caption{GFMC results}
\label{tab:gfmc}
\begin{ruledtabular}
\begin{tabular}{l d d d d}
  $J^\pi$ & 
  \multicolumn{1}{c}{$E$ [MeV]} & 
  \multicolumn{1}{c}{$r_p$ [fm]} &
  \multicolumn{1}{c}{$Q$ [fm$^2$]} &
  \multicolumn{1}{c}{$B(E2\downarrow)$} \\
  \hline
  $0^+$ & -56.3(2) & 2.40    &   0      &  \\
  $2^+$ & -53.1(1) & 2.45(1) &  -9.1(2) & 20.0(8) \\
  $4^+$ & -45.1(2) & 2.48(2) & -12.0(3) & 27.2(15)
\end{tabular}
\end{ruledtabular}
\end{table}
 
A comparison with the {\it ab initio} calculation needs the present experimental result
to be expressed in terms of the B(E2) value for the 4$^+$ to 2$^+$ transition. 
Whereas this is not straightforward, an approximate value can be calculated assuming 
a Breit Wigner form factor for the 4$^+$ resonance and using the 
experimental cross section at the resonance energy. This gives a partial gamma 
width $\Gamma_\gamma$=(0.48$\pm$0.05)~eV
and a B(E2) value of 21$\pm$2.3~$e^2 fm^4$.
This is somewhat lower than the calculated value. However, a better comparison will be possible after 
performing the {\it ab initio} calculation as a function of the alpha particle energy.
The present experimental results, besides putting the $\alpha$-cluster structure of $^8$Be on a firmer 
footing, will provide data for testing the future calculations incorporating the reaction and the structure
aspects in a seamless manner. 

\section {acknowledgement}

We thank the Pelletron crew for delivering the $^4$He beam and R. Kujur and M. Pose for their
help during the experiment.
SP and RBW wish to thank S. C. Pieper for valuable discussions.
The work of CJL and RBW is supported by the US DOE Office of Nuclear Physics 
under Contract No. DE-AC02-06CH11357; the work of SP is supported by the
US NSF under Grant No. PHY-1068305.

\end{document}